\documentclass[aps,prl,twocolumn,superscriptaddress,amssymb,amsmath,nofootinbib]{revtex4-1}

\usepackage{amsmath,setspace,subfigure,amsfonts,latexsym}
\usepackage{amssymb}
\usepackage{color}
\usepackage{epsfig}
\usepackage{graphics}
\usepackage{sidecap}
\definecolor{White}{rgb}{1,1,1}
\definecolor{Red}{rgb}{1,0.1,0}
\definecolor{LightYellow}{rgb}{1,1,.875}
\definecolor{SteelBlue}{rgb}{.273,.508,.703}
\definecolor{navy}{rgb}{0,0,.5}
\definecolor{LightCyan}{rgb}{.875,1,1}
\definecolor{DarkRed}{rgb}{.543,0,0}
\definecolor{HotPink}{rgb}{1,.41,.70}
\definecolor{ForestGreen}{rgb}{.13,.54,.13}
\definecolor{OliveDrab}{rgb}{.42,.55,.14}
\definecolor{MediumBlue}{rgb}{0,0,.80}
\definecolor{RoyalBlue}{rgb}{.25,.41,.88}
\definecolor{DeepSkyBlue}{rgb}{0,.746,1}
\definecolor{Brown}{rgb}{0.545,0.271,0.074}

\def\bea{\begin{eqnarray}}
\def\eea{\end{eqnarray}}
\def\bec{\begin{center}}
\def\ec{\end{center}}

\def\beq{\begin{equation}}
\def\eeq{\end{equation}}

\newcommand\lsim{\mathrel{\rlap{\lower4pt\hbox{\hskip1pt$\sim$}}
    \raise1pt\hbox{$<$}}}
\newcommand\gsim{\mathrel{\rlap{\lower4pt\hbox{\hskip1pt$\sim$}}
    \raise1pt\hbox{$>$}}}
\def\bea{\begin{eqnarray}}
\def\eea{\end{eqnarray}}
\def\ba{\begin{array}}
\def\ea{\end{array}}
\def\bc{\begin{center}}
\def\ec{\end{center}}
\def\nn{\nonumber}

\def\b{\beta}

\def\l{\lambda}

\begin{document}

\title{\Large Coleman-Weinberg Higgs}

\author{Dongjin Chway}
\email{djchway@gmail.com}
\affiliation{Department of Physics and Astronomy
and Center for Theoretical Physics, Seoul National University, Seoul 151-747, Korea}

\author{Radovan Derm\' \i\v sek}
\email{dermisek@indiana.edu}
\affiliation{Physics Department, Indiana University, Bloomington, IN 47405, USA}

\author{Tae Hyun Jung}
\email{thjung0720@gmail.com}
\affiliation{Department of Physics and Astronomy
and Center for Theoretical Physics, Seoul National University, Seoul 151-747, Korea}

\author{Hyung Do Kim}
\email{hdkim@phya.snu.ac.kr}
\affiliation{Department of Physics and Astronomy
and Center for Theoretical Physics, Seoul National University, Seoul 151-747, Korea}

\begin{abstract}

We discuss an extension of the standard model by fields not charged under standard model gauge symmetry in which the 
electroweak symmetry breaking is driven by the Higgs quartic coupling itself without the need for a negative mass term in the potential. This is achieved by a scalar field $S$ with a large coupling to the Higgs field at the electroweak  scale which is driven to very small values at high energies by the gauge coupling of a hidden symmetry under which $S$ is charged. This model can remain perturbative all the way to the Planck scale. The Higgs boson is fully SM-like in its couplings to fermions and gauge bosons. 
However, the effective cubic and quartic self-couplings of the Higgs boson are significantly enhanced.

\end{abstract}

\maketitle

\noindent
{\bf Introduction.}
With the  discovery of the Higgs boson at the Large Hadron Collider (LHC)~\cite{Aad:2012tfa, Chatrchyan:2012ufa}, all the couplings in the standard model of particle physics (SM) are now known. The two parameters of the Higgs potential
with negative quadratic term that destabilizes the potential at the origin, triggering the electroweak symmetry breaking, and the positive quartic term that stabilizes the potential at large scales, 
\bea
V(H) & = & m^2  \; H^\dagger H + \l_h \; (H^\dagger H)^2,
\label{eq:VSM}
\eea
$m^2 < 0$, can be determined from the vacuum expectation value (vev) of the Higgs field, $v \simeq 246$ GeV, previously known from masses of $W$ and $Z$ bosons, and the Higgs boson mass $m_h \simeq 125$ GeV, through well known relations: the condition for the minimum, $-m^2=\l_h v^2$, and the formula for physical Higgs boson mass,  $m_h^2  = 2 \l_h v^2$.
In addition, the Higgs quartic coupling inferred from the measurement of the Higgs boson mass,   $\l^{(2)}_{\rm SM} = \lambda_h$, also determines the strength of self-interactions of the Higgs boson in the SM, namely the cubic, $\l^{(3)}_{\rm SM}$, and quartic, $\l^{(4)}_{\rm SM}$, Higgs couplings, 
\bea
 \l^{(2)}_{\rm SM} & = & \l^{(3)}_{\rm SM} = \l^{(4)}_{\rm SM}.
 \label{selfSM}
\eea 
However confirming this prediction, and gaining confidence that the electroweak symmetry breaking is indeed described by the potential in Eq.~(\ref{eq:VSM}), will be very challenging at the LHC.
Measuring $\l^{(3)}_{\rm SM}$ with 20\%  precision might
  require a linear collider \cite{Peskin:2012we, Baer:2013cma}, and there are curently no prospects to measure $\l^{(4)}_{\rm SM}$. Models for physics beyond the SM typically do not violate relation (\ref{selfSM}) by more than 25\% \cite{Gupta:2013zza}.

A simple alternative to negative mass term triggering spontaneous symmetry breaking, proposed long time ago by Coleman and Weinberg, is the idea that  radiative corrections to the quartic term destabilize the Higgs potential at the origin~\cite{Coleman:1973jx}. In the Coleman-Weinberg (CW) mechanism, the mass term is set to zero, and thus  the model has one less parameter and is classically scale invariant. The quartic coupling of a scalar field, starting from positive value at high energies, is driven through renormalization group (RG) evolution to negative values at low energies. However, in the SM due to large top quark Yukawa coupling, the RG running of the Higgs quartic coupling is exactly opposite and the CW mechanism cannot be realized. It has also been difficult to realize the CW mechanism for $\l_{h}$ in extensions of the SM.  Large couplings of new fields to the Higgs boson must exist in a given extension, and this makes the theory non-perturbative already at or very near the electroweak(EW) scale~\cite{Foot:2007as, AlexanderNunneley:2010nw}.
There is no example of a model in which the EW symmetry breaking is fully triggered by radiative corrections to $\l_h$ which would remain perturbative significantly above the EW scale. 
 
In this letter we provide the first such model which can remain perturbative all the way to the Planck scale. 
We show that, in a simple extension of the standard model by fields not charged under standard model gauge symmetry, the EW symmetry breaking can be driven by the Higgs quartic coupling itself without the need for a negative mass term in the potential. This is achieved by a scalar field $S$ with a large (but perturbative) coupling to the Higgs field at the EW scale which is driven to very small values at high energies by the gauge coupling of a hidden symmetry under which $S$  is charged. The Higgs boson is fully SM-like in its couplings to fermions and gauge bosons. The  effective cubic and quartic self-couplings of the Higgs boson are, however, dramatically modified from SM predictions:
\bea
\l^{(3)}_{\rm CW} & = & \frac{5}{3} \l^{(2)}_{\rm CW} , \hspace{5mm}
\l^{(4)}_{\rm CW} = \frac{11}{3} \l^{(2)}_{\rm CW},
\label{selfCW}
\eea
in the leading order and are subject to potentially significant corrections.
We call  the Higgs boson with these properties {\it Coleman-Weinberg Higgs}.

The scenario we discuss is unique and easily distinguishable from singlet extensions of the SM in which the CW mechanism is utilized to generate a vev of the extra singlet which directly participates in EW symmetry breaking~\cite{Hempfling:1996ht, Meissner:2006zh, Hambye:2007vf}. For recent discussions in a variety of contexts, see also Refs.~\cite{Englert:2013gz, Heikinheimo:2013fta, Hambye:2013dgv}. In our model, the extra singlets are merely spectators. They do not get vevs and do not directly participate in EW symmetry breaking.

Setting the mass term to zero and working in the unitary gauge,  $H^T(x)=1/\sqrt{2} (0,\phi(x))$, 
the potential can be written in terms of  real scalar field $\phi$ as $V_{\text{eff}}(\phi) = \frac{1}{4}(\l_h (\mu=a \phi)+\delta \l_h) \phi^4 e^{4\Gamma(a\phi)}$, 
where the $a$ is a function of couplings, $\delta \l_h(\mu)$ is the radiative correction (both will be given after specifying the model) and $\Gamma(\mu)=\int^{\mu}_{M_Z}d \ln\mu' \ \gamma(\mu')$ with $\gamma$ the Higgs field anomalous dimension.
The condition for the minimum in terms of  $\hat{\l}_h \equiv (\l_h+\delta \l_h)e^{4\Gamma}$ is
\begin{equation}
\left. \frac{d V_{\text{eff}}}{d\phi} \right|_{\phi=v} =  (\hat{\l}_h + \frac{\hat{\l}'_h}{4}) \left. \phi^3 \right|_{\phi=v} = 0 \ \to  \ \hat{\l}_h+\frac{\hat{\l}'_h}{4}=0, 
\label{eq:min}
\end{equation}
where $\hat{\l}'_h = \phi d\hat{\l}_h/d\phi $. 
The physical Higgs mass originates from the second derivative of the potential,
\bea
m_h^2 &=& \left. \frac{d^2 V_\text{eff}}{d\phi^2}\right|_{\phi=v} = \left.(3\hat{\l}_h+\frac{7}{4}\hat{\l}'_h+\frac{1}{4}\hat{\l}''_h)\phi^2 \right|_{\phi=v}    \nn \\
&=& (-4\hat{\l}_h+\frac{1}{4}\hat{\l}''_h) v^2 = (\hat{\l}'_h +\frac{1}{4}\hat{\l}''_h)v^2 , \label{eq:mh} 
\eea
and it receives the momentum dependent corrections that we include in the numerical results.
In this letter we keep all the terms coming from derivatives of the effective potential so that the equations are valid beyond one loop.

In analogy with the SM, we define the coupling
\begin{equation}
\l^{(2)}_{\rm CW} \equiv  \left.\frac{1}{2v^2}\frac{d^2V_\text{eff}}{d\phi^2}\right|_{\phi=v} = \frac{1}{2}( \hat{\l}'_h+\frac{\hat{\l}''_h}{4}). \label{l2CW}
\end{equation}
Since it is determined from the physical Higgs mass, numerically $\l^{(2)}_{\rm CW} \simeq \l^{(2)}_{\rm SM}$ (equal up to momentum dependent corrections).
To realize the CW mechanism without any other field getting a vev, the  observed Higgs boson mass requires small negative quartic coupling, $\hat{\l}_h \simeq -1/16$ and $\hat{\l}'_h \simeq 1/4$, at the minimum of the potential. 

\noindent
{\bf The Model.} We introduce a complex scalar, $S$,  in the fundamental representation of extra $SU(N_S)$ gauge symmetry.
The general form of the classically scale invariant scalar potential is given by
\bea
V = \l_h (H^\dagger H)^2 + \l_{hs} (H^\dagger H)(S^\dagger S) + \l_s (S^\dagger S)^2,
\label{eq:V}
\eea
and the  RG equations for quartic couplings are as follows:
{\small
\bea
16\pi^2 \frac{d\l_h}{dt} &&= 24\l_h^2 + N_S\l_{hs}^2-6y_t^4+12y_t^2\l_h
\\
16\pi^2 \frac{d\l_s}{dt} &&= 4(4+N_S)\lambda_s^2+2\lambda_{hs}^2 -6\Big(\frac{N_S^2-1}{N_S}\Big) g_4^2 \lambda_s\nn\\
&& +\frac{3}{4}\Big(\frac{N_S^3+N_S^2-4N_S+2}{N_S^2} \Big)g_4^4, \label{ls} \\
16\pi^2 \frac{d\l_{hs}}{dt} &&= \l_{hs}\Big[4\l_{hs}+12\l_h\nn \\
&&+(4N_S+4)\l_s  -3\Big(\frac{N_S^2-1}{N_S}\Big) g_4^2+6y_t^2 \Big],  \label{eq:b_l_hs}
\eea
}
where $y_t$ is the top Yukawa coupling, $g_3$ is the gauge coupling of $SU(3)_c$, $g_4$ is the gauge coupling of $SU(N_S)$ with the RG equation
{\small
\bea
16\pi^2 \frac{d g_4}{dt} &&=-g_4^3\big(\frac{11}{3}N_S-\frac{2}{3}N_f-\frac{1}{6}\big),
\eea}
where $N_f$ is the number of Dirac fermions in the fundamental representation of $SU(N_S)$.
For $y_t$ and $g_3$, the RG equations are the same as in the SM.
For simplicity, we do not write the contributions from SM gauge couplings of $SU(2)_L\times U(1)_Y$ which are negligible.

The extra singlet does not get a vev. Its sole purpose is to generate large enough positive contribution to the $\b_{\l_h}$ through the mixed quartic coupling $\l_{hs}$. Thus the conditions for the minimum (\ref{eq:min}) and the formula for the Higgs mass (\ref{eq:mh}) are unchanged, and the measured value of the Higgs boson mass is obtained for $N_S \l_{hs}^2 \simeq 40$.
The  singlet scalar gets the mass from the electroweak symmetry breaking and it is fixed by $N_S$,
{\small
\bea
m_S & = & \sqrt{\l_{hs}} \frac{v}{\sqrt{2}} \;  \simeq \; \frac{440}{N_S^{1/4}} \ {\rm GeV}.
\label{eq:mS}
\eea
}
The anomalous dimension $\gamma$ is that of the SM at one loop and 
the threshold correction $\delta \l_h$ is given by
{\small
\begin{equation}
\delta \l_h=\frac{1}{2}\frac{N_S \l_{hs}^2}{16\pi^2}\Big(\ln\big(\frac{m_S^2}{\mu^2}\big)-\frac{3}{2}+2\Gamma \Big)+({\rm{ SM \ part}}).
\end{equation}}
Because $\l_{hs}$ is the largest coupling, we choose $\mu=m_S$, i.e. $a^2=\l_{hs}/2$, to eliminate the logarithmic contribution. 
Once we choose $a$, we can obtain $\hat{\lambda}_h^\prime$ in terms of couplings $x_i$ and their beta functions $\beta_{x_i} = d{x_i}/d\ln{\mu}$,
{\small
\begin{equation}
\hat{\lambda}_h^\prime (x_1,\cdots,x_n;\mu=a\phi)=\frac{1}{1-\frac{\beta_a}{a}}\sum_i \beta_{x_i}\frac{\partial}{\partial x_i} \hat{\lambda}_h.
\end{equation}
}
Although our choice of $\mu$ removes the largest logarithmic corrections appearing 
in higher orders, it is not unique and we will also discuss the $\mu$ dependence of the results.

\begin{figure}[t]
\includegraphics[width=0.4\textwidth]{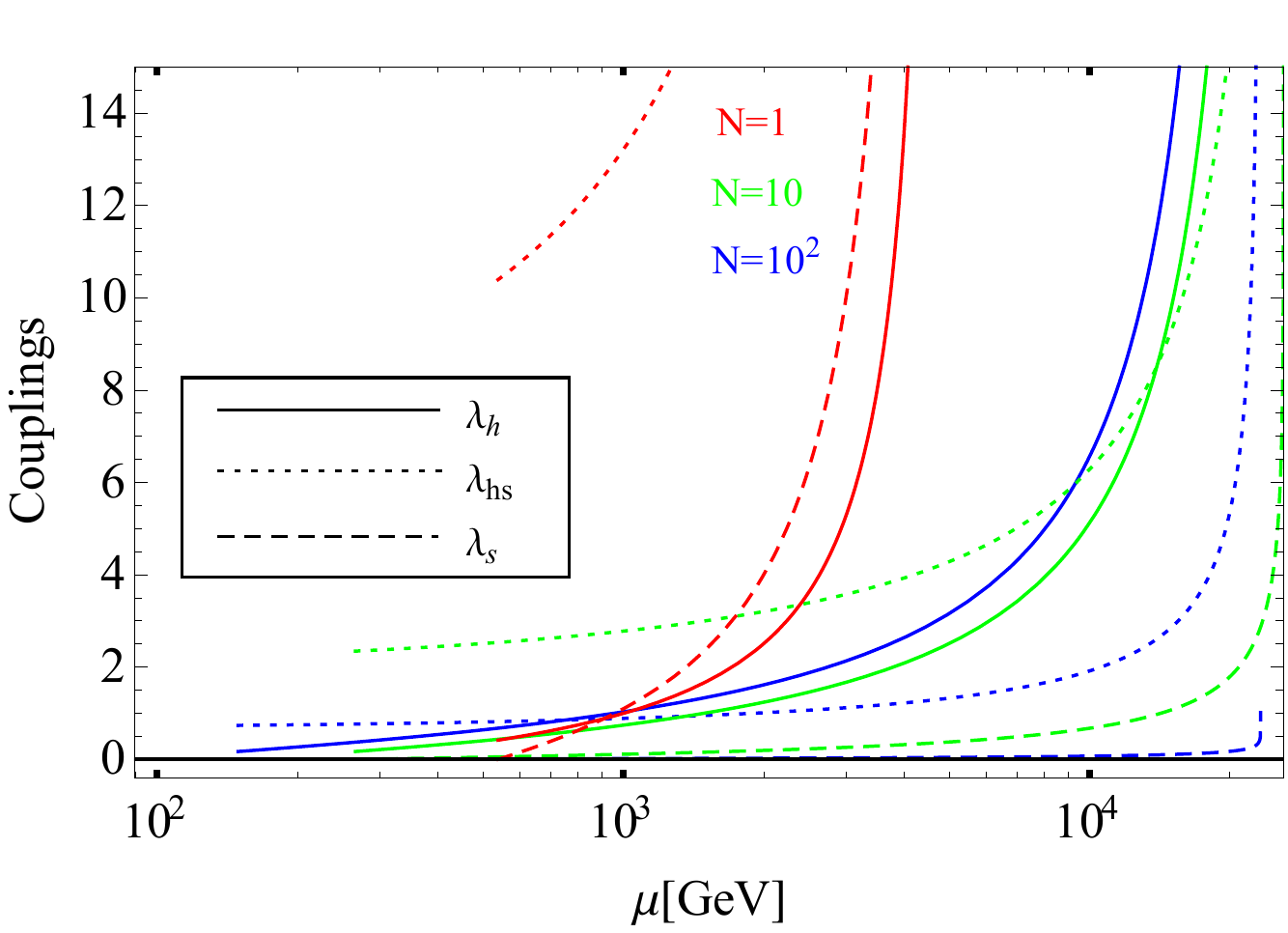}
\caption{The RG  evolution of quartic couplings $\lambda_{h,s,hs}$ in the model with $N_S= 1$ (red), 10 (green) and $10^2$ (blue) complex scalars for $g_4=0$. The effective potential is minimized at $\mu = \max (m_S, m_h)$ which depends on $N_S$. Boundary conditions at this scale for $\lambda_h$ and $\lambda_{hs}$ are determined by the observed Higgs boson mass, $\hat{\lambda}_h (m_S) \simeq -1/16$, and $\lambda_s$ is set to 0 for simplicity.}
\label{basic}
\end{figure}

To illustrate the importance of the extra gauge interactions, let us first  discuss the case without the gauge symmetry by setting $g_4=0$.
For simplicity, we  keep  $SU(N_S)$ global symmetry, namely couplings of all $N_S$ complex scalars are assumed to be identical. 
The RG evolution of quartic couplings
for 3 different $N_S$ is shown in Fig.~\ref{basic}. 
For one extra complex scalar, to obtain the observed value of the Higgs boson mass, extremely large mixed quartic coupling is required. This coupling blows up rapidly just above the TeV scale. With increasing $N_S$, smaller $\l_{hs}$ is needed for the observed Higgs boson mass, and thus the model remains perturbative up to higher energies. However, large $\beta_{\l_h} $ implies that $\l_{h}$  runs fast, and for $N_S = 10$, it is this coupling that blows up first, at about 20 TeV. This feature is fully fixed by the requirement   $\hat{\l}'_h \simeq 1/4$ and does not depend on $N_S$. Thus the perturbativity cannot be extended beyond 20 TeV
by increasing $N_S$, which is illustrated by the $N_S = 10^2$ case. These conclusions do not depend on the assumption of equal couplings of all scalars.

\begin{figure}[t]
\includegraphics[width=0.4\textwidth]{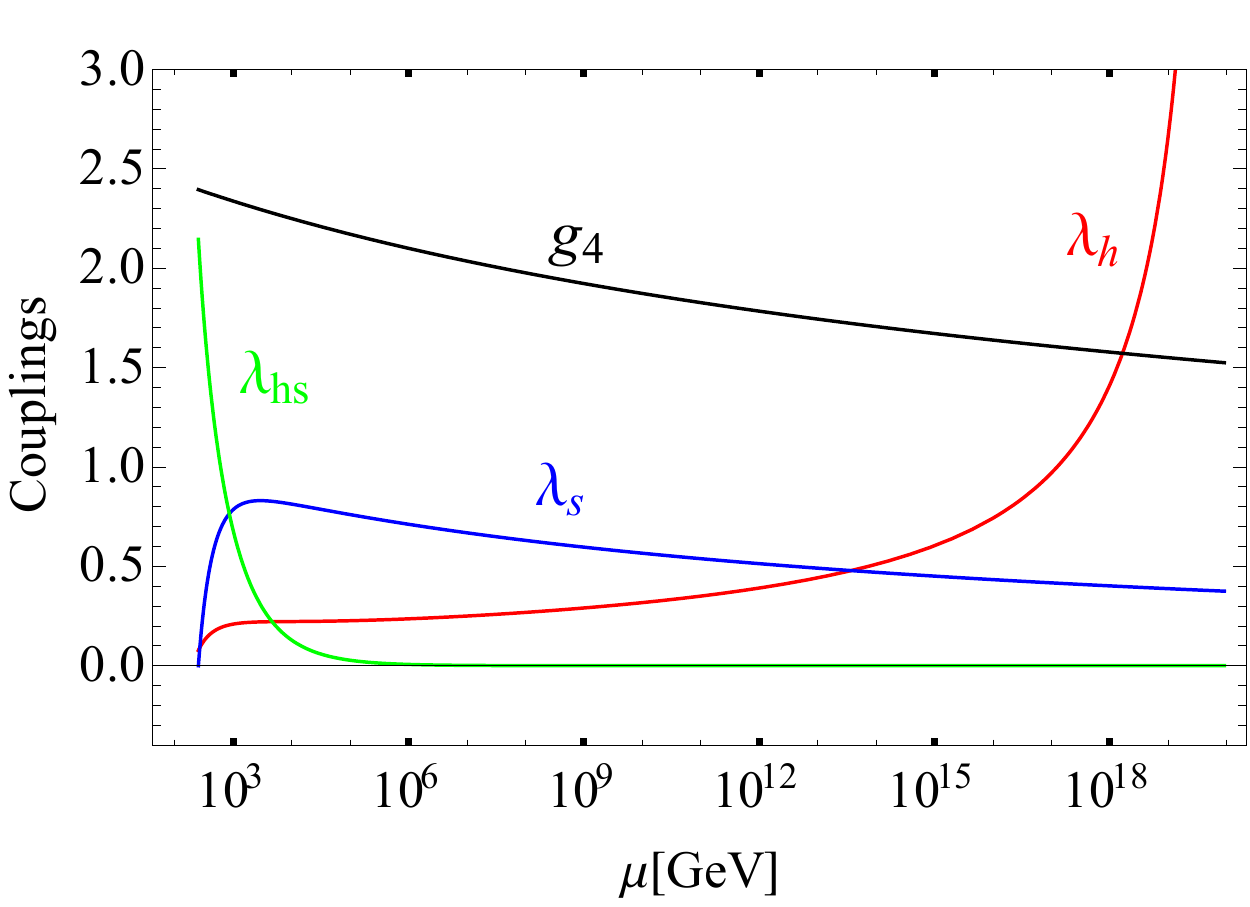}
\caption{The RG evolution of quartic couplings and extra gauge coupling in the model with  $N_S=10$. The effective potential is minimized at $\mu = m_S$. Boundary conditions at this scale for $\lambda_h$ and $\lambda_{hs}$ are determined by the observed Higgs boson mass, $\hat{\lambda}_h (m_S) \simeq -1/16$, and $\lambda_s$ is set to 0 for simplicity.}
\label{extgauge}
\end{figure}

The perturbativity can be extended if  $\l_{hs}$ drops quickly above the EW scale. This is achieved by sizeable gauge coupling $g_4$ of the hidden $SU(N_S)$, see Eq.~(\ref{eq:b_l_hs}).
The rapid decrease of $\l_{hs}$  slows down the running of $\l_h$ above the EW scale. At the same time, there is a quasi-fixed point  for $\l_{s}/g_4^2$ in the UV which, if
 the running of $g_4$ is not significant, is expected from Eq.~(\ref{ls}) for $N_S \ge 4$.
The evolution of couplings in the model with $N_S=10$ is shown in Fig.~\ref{extgauge}. 
All the couplings remain perturbative up to the Planck scale.
 
Slow running of $g_4$ (which decreases $\l_{hs}$ exponentially) can be achieved by adjusting the number of SU($N_S$) charged fermions with no couplings to S (in our example  we assume 54 spectator fermions in $10$ and $\overline{10}$ of SU(10)). In order to avoid extra light states, either for cosmological reasons or to avoid new decay modes of the Higgs boson, we for simplicity assume that the extra fermions acquire masses near the $m_S$ scale, see Eq.~(\ref{eq:mS}). This can be always achieved without violating classical conformality and without affecting the properties of the Higgs boson. Below the scale  of extra fermions the running of $g_4$ is accelerated and  confinement occurs very fast.

An interesting insight can be gained by looking at the evolution in Fig.~\ref{extgauge} starting from the UV scale. The $\l_{hs}$ is tiny in the UV and so the two sectors evolve almost separately: $\lambda_s$ follows $g_4$ and $\l_{h}$ runs down due to itself. This can be stretched over many orders of magnitude in the energy scale. However, eventually, $\l_{hs}$ becomes  sizable and provides large positive contribution to $\b_{\l_{h}}$. Thus, instead of turning back up  due to  top Yukawa coupling near the EW scale,  $\l_{h}$ is driven to negative value which triggers EW symmetry breaking.

\begin{figure}[t]
\includegraphics[width=0.35\textwidth]{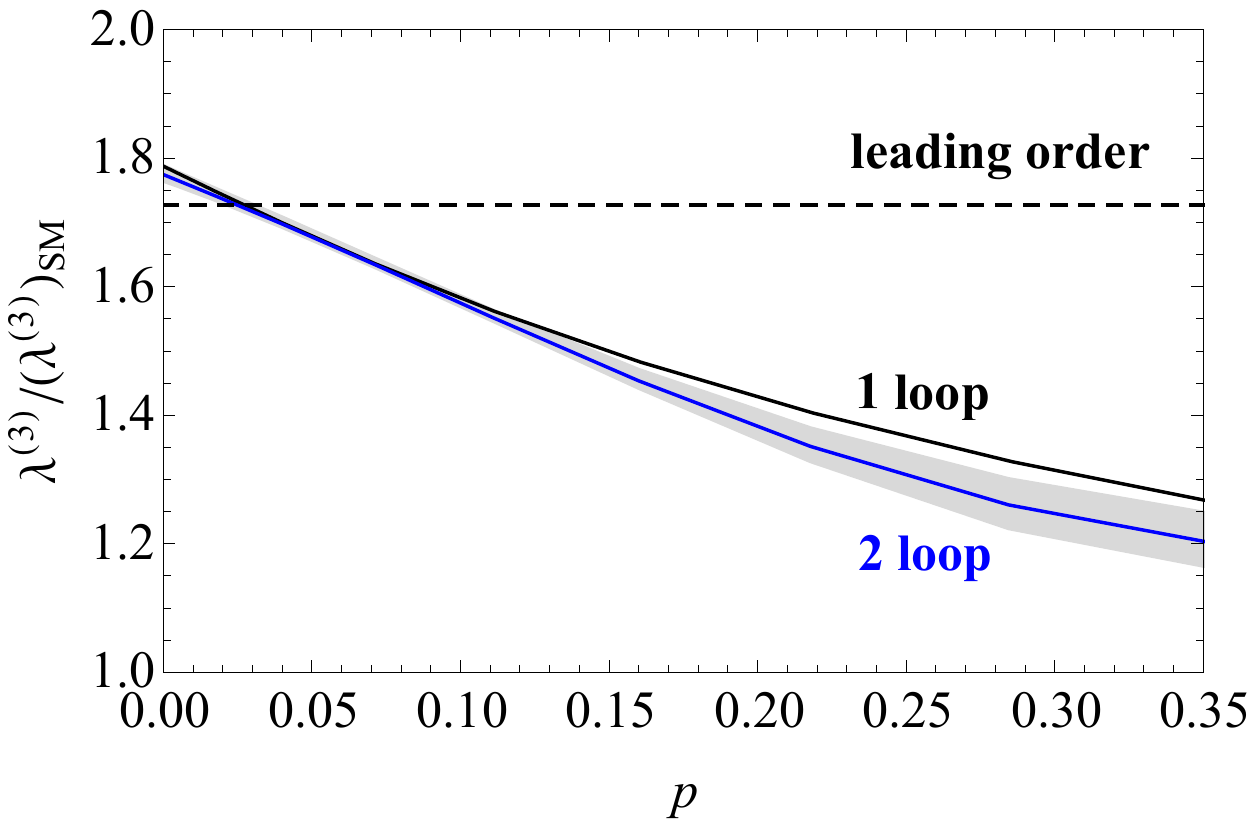}
\includegraphics[width=0.35\textwidth]{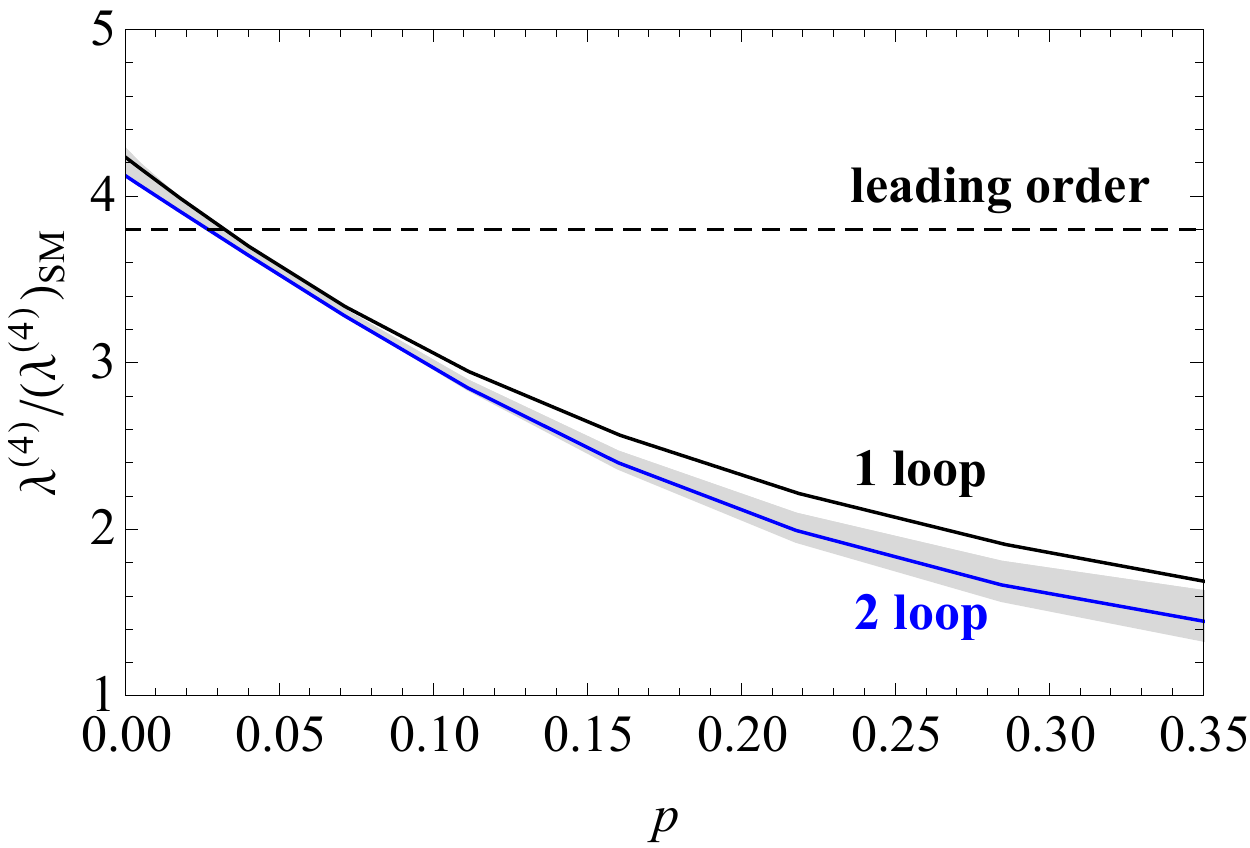}
\caption{Ratios of predicted effecive cubic and quartic couplings and the SM value versus $p$, where $p =N_S g_4^2/(16\pi^2)$ and $N_S=10$. 
Dashed line indicates leading order predictions that includes the momentum dependent corrections to the Higgs pole mass.
Solid curves represent predicted values at one and two loop keeping all the terms coming from derivatives of the effective potential. Gray band shows the $\mu$ dependence of the two loop predictions. We vary $\mu$ from $m_S/\sqrt{2}$ to $\sqrt{2} m_S$.}
\label{self3}
\end{figure}

\noindent
{\bf Self-couplings of the Higgs boson.} If the EW symmetry breaking is fully achieved by negative $\hat{\l}_h$,
the effective Higgs cubic and quartic couplings, defined in analogy with the SM,  are
{\small
\bea
\l^{(3)}_{\rm CW} & \equiv & \left.\frac{1}{6v} \frac{d^3V_{\text{eff}}}{d\phi^3}\right|_{\phi=v}
= \frac{5 \l^{(2)}_{\rm CW}}{3} +\frac{\hat{\l}''_h}{6}+\frac{\hat{\l}'''_h}{24}, \\ 
\l^{(4)}_{\rm CW} & \equiv  & \left.\frac{1}{6} \frac{d^4V_{\text{eff}}}{d\phi^4}\right|_{\phi=v}
= \frac{11\l^{(2)}_{\rm CW}}{3}  +\hat{\l}''_h+\frac{5\hat{\l}'''_h}{12}+\frac{\hat{\l}''''_h}{24},
\eea
}
where the condition for the minimum (\ref{eq:min}), Higgs mass (\ref{eq:mh}) and Eq.~(\ref{l2CW}) are plugged in. 
If we neglect $\hat{\l}''_h$ and higher derivatives, these effective couplings are enhanced  in the leading order by 67\%  and 267\%  respectively compared to  the SM prediction (\ref{selfSM}), as already presented in Eq.~(\ref{selfCW}).
 Actually, the momentum dependent correction to the Higgs pole mass,  not explicitly shown in (\ref{eq:mh}), makes the effective couplings slightly larger as the dashed lines indicate in Fig.~\ref{self3}.

The $\hat{\l}''_h$ depends on large $g_4$ coupling and thus the corrections to the leading order predictions for the effective cubic and quartic couplings are sizeable (the effects of third and fourth derivatives of $\hat{\l}_h$ are however numerically small). Keeping all the derivatives, the Higgs self couplings as functions of $g_4^2$ are given in Fig.~\ref{self3}. We show results both at one loop, based on the formulas presented in this letter, and two loop, based on the two loop renormalisation group equations and threshold corrections derived in Ref.~\cite{Chway:2014}.  For the two loop results we also show the dependance on the choice of $\mu$ that indicates the size of higher loop corrections. The example in Fig.~\ref{extgauge} which extends perturbativity up to the Planck scale corresponds to $p=0.35$ in Fig.~\ref{self3}. Relaxing this requirement allows for smaller values of the $g_4$. As $g_4$ decreases, the predicted values of   Higgs self couplings are getting larger and closer to the leading order prediction given in Eq.~(\ref{selfCW}). 

\noindent
{\bf Discussion and Conclusions.}
In the Coleman-Weinberg Higgs scenario that we discussed, the EW symmetry breaking is fully achieved by the Higgs quartic coupling being driven to negative values at the EW scale. 
This is achieved by a scalar field $S$, singlet under SM gauge symmetry, but charged under extra gauge symmetry $SU(N_S)$ which keeps the model perturbative to the Planck scale.
Since the Higgs doublet is the only field getting a vev, the Higgs boson is fully SM-like in its couplings to fermions and gauge bosons. 
However, the effective cubic and quartic self-couplings of the Higgs boson are significantly enhanced. Depending on the size of hidden gauge coupling $g_4$ which determines the scale up to which the theory can remain perturbative, the cubic coupling is enhanced by about 20\% (perturbative to Planck scale) to 80\% (perturbative to 10 TeV) and quartic coupling by 40\% to 300\%.

Interestingly, a number of extra  scalar fields with couplings to the Higgs field, similar to our scenario,  have been introduced to provide strong first order electroweak phase transition which can make the electroweak baryogenesis viable~\cite{Espinosa:2007qk, Noble:2007kk, Espinosa:2008kw, Chung:2012vg}. 

Although extra scalars are stable, their relic density  from thermal freeze out  is suppressed due to large extra gauge coupling.
They contribute negligibly  to the dark matter of the Universe, and expected signals are far below sensitivities of near future direct detection experiments. 

Since the mass of extra scalars is $440/N_S^{1/4}$ GeV,  they can be produced at the LHC and constrained by  monojet searches. However, extra scalars only couple to the Higgs boson and thus their production cross section (through an off-shell Higgs boson produced in gluon fusion or in association with a vector boson) is suppressed. We leave the study of LHC signatures for future work.

\noindent
{\bf Acknowledgements}
R. D. and H.D. K. thank the Galileo Galilei Institute for Theoretical Physics for the hospitality and the INFN for partial support during the completion of this work. The work of H.D. K. and T. J. was supported  by the NRF of Korea No. 2011-0017051. The work of R. D. was supported in part by the United States Department of Energy under Grant No. DE-FG02-91ER40661.

\end{document}